\documentclass[prb,twocolumn,superscriptaddress,showpacs,
amsmath,amssymb]{revtex4}
\usepackage{graphicx} 
\usepackage{dcolumn} 

\begin{document}

\title{The imperfect Bose gas in $d$ dimensions: critical behavior and Casimir forces}

\author{M. Napi\'{o}rkowski}
\author{P. Jakubczyk}
\email{pjak@fuw.edu.pl} 
\author{K. Nowak}
\affiliation{Institute of Theoretical Physics, Faculty of Physics, University of Warsaw, 
 Ho\.za 69, 00-681 Warsaw, Poland}

\date{\today}

\begin{abstract}
We consider the $d$-dimensional imperfect (mean-field) Bose gas confined in a slit-like geometry and subject to periodic boundary conditions. Within an exact analytical treatment we first extract the bulk 
critical 
properties of the system at Bose-Einstein condensation and identify the bulk universality class to be the one of the classical $d$-dimensional spherical model. Subsequently we consider finite slit 
width $D$ and analyze the excess surface free energy and the related Casimir force acting between the slit boundaries. Above the bulk condensation temperature ($T>T_c$) the Casimir force decays 
exponentially as a function of $D$ with the bulk correlation length determining the relevant length scale. For $T=T_c$ and for $T<T_c$ its decay is algebraic. The magnitude of the Casimir forces at $T_c$ 
and for $T<T_c$ is governed by the universal Casimir amplitudes. We extract the relevant values for different $d$ and compute the scaling functions describing the crossover between the critical and 
low-temperature asymptotics of the Casimir force. The scaling function is monotonous at any $d\in (2,4)$ and becomes constant for $d>4$ and $T\leq T_c$.

\end{abstract}

\maketitle

\section{Introduction}
The non-interacting Bose gas has played a prominent role in the historical development of the quantum many-body physics and is among the most basic and earliest studied model systems featuring a phase 
transition.\cite{Ziff_77,Bal07} It also serves these days as the most standard textbook example illustrating the phenomenon of Bose-Einstein condensation. 
Despite this indisputable historical and pedagogical role, it is of limited use for quantitative comparison to realistic systems 
due to the indispensable role of interactions. It is also not quite satisfactory for fundamental reasons rooted in the lack of superstability 
\cite{Ziff_77, Buffet_83}  leading to inequivalence between descriptions of certain features within 
different Gibbs ensembles. The thermodynamics of the ideal Bose gas is defined only for nonpositive values of the chemical potential $\mu$ and the condensate 
occurs only at $\mu=0$ for $T < T_{c}$.    

Some of these pathologies are cured when the ideal Bose gas is supplemented with a mean-field-like interparticle interaction resulting in a model known as the 
imperfect Bose gas (IBG). 
\cite{Davies_72, Zagrebnov_01, Lewis_book, Berg_84} In particular, the thermodynamics of the IBG is defined for arbitrary values of $\mu$ and the region of the  $\mu - T$ phase diagram in which the condensate occurs corresponds to $\mu > \mu_{c}(T) > 0$. 
The IBG remains susceptible to mathematically rigorous analysis and its physical content is clarified by a formulation of the mean-field theory with the help of the repulsive binary potential subject to 
the Kac scaling.\cite{Lewis_book} As turns out, the IBG corresponds to the Kac limit, where the binary interaction becomes progressively weaker and long-ranged. 

The bulk thermodynamics of the three-dimensional imperfect Bose gas was studied rigorously decades ago.\cite{Davies_72} A convenient formulation of the problem was recently proposed in 
Ref.~\onlinecite{Napiorkowski_11}, where a finite-size effect leading to the thermodynamic Casimir force was addressed in addition to reproducing the known results in the thermodynamic limit. The present 
work supplements and extends Ref.~\onlinecite{Napiorkowski_11} by considering the system at arbitrary spatial dimensionality $d>2$ and additionally addressing the crossover between the critical Casimir 
forces caused by the soft fluctuations at $T_c$, and the long-ranged (Goldstone-type) modes below $T_c$. This crossover is known to be governed by a universal scaling function, whose shape we study for 
different values of $d$. The paper is structured as follows: 

In Sec.~II we introduce the model and the steepest descent method which we will use in the calculations. The method is adapted from Ref.~\onlinecite{Napiorkowski_11}. 
In Sec.~III we consider the system in the 
thermodynamic limit, where we extract the bulk phase diagram and the values of the correlation length and specific heat critical exponents ($\nu$ and $\alpha$, respectively) as a function of the system spatial dimensionality $d$. 
As turns out, the obtained dependencies $\nu(d)$ and $\alpha(d)$ are the 
same as for the classical $d$-dimensional spherical model. In Sec.~IV we consider the system to be of finite extension $D$ in one direction and calculate the excess grand canonical free energy, 
focusing on the case of macroscopically large separations $D$. We extract the universal scaling function governing the crossover between the excess grand canonical free energy at the bulk transition 
temperature $T_c$ and below it. We investigate the evolution of the scaling function upon varying dimensionality $d$ and find that the function attains a trivial (constant) shape above $d=4$. We also 
compute the scaling function above $T_c$. In Sec.~V we discuss the closely related Casimir forces and make a comparison to earlier results addressing the classical spherical model. We summarize the work 
in Sec.~VI.

 \section{Imperfect Bose gas and the stationary point method}
In this section we present the model and give a representation of its grand-canonical partition function following Ref.~\onlinecite{Napiorkowski_11} but keeping the dimensionality $d$ arbitrary. 
Thus the reasoning contained in the earlier work is adapted here to the case of $d$ being a continuously variable parameter. The Hamiltonian of the IBG\cite{Davies_72} reads 
 \begin{equation}
  \hat{H}_{IBG}=\sum_{\bf k} \frac{\hbar^2{\bf k}^2}{2m}\hat{n}_{\bf k}+\frac{a}{2V}\hat{N}^2 \;,
 \end{equation}
 where we use the standard second-quantized notation. The Hamiltonian involves the quadratic kinetic term and the mean-field interaction energy.  
The latter, $H_{mf}=\frac{a}{2V}\hat{N}^2$ ($a>0$) may be obtained 
from a Hamiltonian involving a binary interaction $v(r)$ in the Kac limit $\lim_{\gamma\to 0}\gamma^dv(\gamma r)$, where the interaction strength vanishes, but its 
range diverges. 
The particles are spinless and the gas is enclosed in a box of volume $V= L^{d-1}\times D$, where $L$ is assumed to be much larger than any other length scale present in the system. The system is subject to periodic 
boundary conditions, implying $k_i=\frac{2\pi n_i}{L}$ for $i\in\{1,...d-1\}$ and $k_d=\frac{2\pi n_d}{D}$. Here $n_i\in\mathbb{Z}$.  
The grand canonical partition function of the IBG can be conveniently represented as\cite{Napiorkowski_11}
\begin{equation}
\label{GC}
 \Xi (T,L,D,\mu) = -i e^{\frac{\beta\mu^2}{2a}V}\left(\frac{V}{2\pi a \beta}\right)^{1/2} \int_{\alpha\beta-i\infty}^{\alpha\beta+i\infty}dse^{-V\phi(s)}\;, 
\end{equation}
where 
\begin{eqnarray}
\label{phi_def}
\phi(s)=-\frac{s^2}{2a\beta}+\frac{s\mu}{a}+\frac{1}{V}\sum_{n_d}\ln(1-e^{s-\beta\epsilon_{k_d}})- \\ \nonumber
\sum_{n_d}\frac{1}{D\lambda^{d-1}}\,\,g_{\frac{d+1}{2}}(e^{s-\beta\epsilon_{k_d}})\;,
\end{eqnarray}
with $\beta\epsilon_{k_d}=\frac{\lambda^2}{D^2}\pi n_d^2$, and $\lambda = h/\sqrt{2 \pi m k_{B} T}$ denoting the thermal wavelength. The Bose functions are defined via
\begin{equation}
 g_n(z) = \sum_{k=1}^{\infty}\frac{z^k}{k^n}\;.
\end{equation}
The parameter $\alpha$ in the contour of integration in Eq.~(\ref{GC}) is negative which ensures the convergence of the series defining the analytic continuation of the perfect gas partition 
function, Ref.~\onlinecite{Napiorkowski_11}. 

For the reference bulk system, where the length scales $L$ and $D$ are of the same order of magnitude and are ultimately considered infinite, we have
\begin{equation}
\label{GC_bulk}
\Xi_b (T,V,\mu) = -i e^{\frac{\beta\mu^2}{2a}V}\left(\frac{V}{2\pi a \beta}\right)^{1/2}\int_{\alpha\beta-i\infty}^{\alpha\beta+i\infty}dse^{-V\phi_b(s)}\;, 
\end{equation}
with
\begin{equation}
\label{phi_b_def}
 \phi_b(s)= -\frac{s^2}{2a\beta}+\frac{\mu s}{a}-\frac{g_{\frac{d}{2}+1}(e^s) }{\lambda^d}+\frac{1}{V}\ln(1-e^s) \;.
\end{equation}
We note that the stationary point approximation to the integrals in Eq.~(\ref{GC}) and in Eq.~(\ref{GC_bulk}) becomes exact in the limit $L\to\infty$, i.e., 
\begin{equation}
\label{GCA}
\frac{1}{V}\ln\Xi(T,L,D,\mu)\longrightarrow \left(\frac{\beta\mu^2}{2a}-\phi(\bar{s})\right)\;,
\end{equation}
where $\bar{s}$ is the stationary point. 
The problem of evaluating the partition function becomes for $L\to\infty$ equivalent to solving the stationary point equation 
\begin{equation}
\label{stat_point_eq}
 \phi '(\bar{s})=0\;, 
\end{equation}
and analogously for the reference bulk system, where $\phi_b(s)$ replaces $\phi(s)$.

 \section{The bulk system}
In this section we consider the bulk system, putting $D\approx L$ and passing to the thermodynamic limit. The stationary-point equation $\phi_b '(s)|_{s=s_0}=0\;$ yields 
\begin{equation}
\label{bulk_sp}
0=\frac{1}{a\beta}(-s_0+\mu\beta)-\frac{g_{\frac{d}{2}}(e^{s_0})}{\lambda^d}-\frac{1}{V}\frac{e^{s_0}}{1-e^{s_0}}   \;, 
\end{equation}
and the grand-canonical free energy density 
\begin{eqnarray}
\omega_b(T,\mu)=-\lim_{V\to\infty}(\beta V)^{-1}\ln\Xi_b(T,V,\mu)=\nonumber \\
-\frac{\mu^2}{2a}+\beta^{-1}\phi_b(s_0).
\end{eqnarray}
For $d\leq 2$ the function 
$g_{\frac{d}{2}}(e^{s_0})$ features a singularity at $s_0=0$, and Eq.~(\ref{bulk_sp})  admits only negative solutions. This results in finite bulk correlation length 
$\xi\sim|s_0|^{-1/2}$ 
(see Refs.~\onlinecite{Napiorkowski_11, Napiorkowski_12}) and the absence of the Bose-Einstein condensation at any $\mu$ and $T>0$. On the other hand, for $d>2$ 
the function $g_{\frac{d}{2}}(e^{s_0})$ is finite for $s_0=0$. Consequently, 
Eq.~(\ref{bulk_sp}) features in the limit $V\to\infty$ a unique solution $s_0<0$ provided $\frac{\lambda^d\mu}{a}<g_{\frac{d}{2}}(1)$. This defines the normal 
(high-temperature) phase of the system. On the other hand, 
for the thermodynamic state chosen so that $\frac{\lambda^d\mu}{a}\geq g_{\frac{d}{2}}(1)$ and $V\to\infty$ one finds from Eq.~(\ref{bulk_sp}) that $|s_0|\to 0$. 
In this case the last term in   
Eq.~(\ref{bulk_sp}) gives a nonzero contribution equal to the condensate density. Thus we identify the critical value of the chemical potential 
\begin{equation}
 \mu_c(T)=g_{\frac{d}{2}}(1)\frac{a}{\lambda^d}=\zeta\left(\frac{d}{2}\right)\frac{a}{\lambda^d},
\end{equation}
above which - at a given temperature - the system displays a phase involving the Bose-Einstein condensate. Here $\zeta(z)$ is the Riemann function. 
The above expression for the critical line $\mu_{c}(T)$ can be rewritten as $\mu_{c}(T)=a\, n_{c}(T)$, where $n_{c}$ is the critical density of the $d$-dimensional ideal Bose gas. 


We now analyze the system for $|s_0|\ll 1 $, corresponding to the thermodynamic state either in the high-temperature phase close to Bose-Einstein condensation, or below the condensation temperature, i.e., in the phase hosting the condensate. The asymptotic behavior of the 
Bose functions in the limit $|s_0|\to 0$ is given by\cite{Ziff_77}
\begin{eqnarray}
\label{g_expansion}
g_{\frac{d}{2}}(e^{s_{0}})\,-\,\zeta\left(\frac{d}{2}\right) \approx \left\{ 
\begin{array}{l l}
  \Gamma(1-\frac{d}{2})\,|s_{0}|^{\frac{d-2}{2}} & \quad \mbox{$2<d<4$} \\
  |s_{0}|\,\log(|s_{0}|) & \quad \mbox{$d=4$} \\ 
	-\,\zeta(\frac{d}{2}-1) \, |s_{0}| & \quad \mbox{$d>4$}  \quad. \\ \end{array}  \right. 
\end{eqnarray}
To shorten the notation we introduce the dimensionless distance from the 	critical line $\epsilon = \frac{ \mu - \mu_{c}}{\mu_{c}}$. 
The relevant parameter determining the magnitude of $s_0$ depends on the sign of $\epsilon$. We analyze the distinct cases separately.

\subsection{The high-temperature phase} 
The high-temperature (normal) phase corresponds to $\epsilon <0$.  We pass to the limit $V\to\infty$ in Eq.~(\ref{bulk_sp}), and for $\epsilon \to 0^{-}$ use the 
asymptotic formulae in Eq.~(\ref{g_expansion}) to find   
\begin{eqnarray}
\label{s0}
|s_{0}(T,\mu)| = \left\{ 
\begin{array}{l l}
  \left(\frac{\zeta(\frac{d}{2})}{|\Gamma(1-\frac{d}{2})|}\right)^{\frac{2}{d-2}}\,\left|\epsilon \right|^{\frac{2}{d-2}} & \quad \mbox{$2<d<4$} \\
  - \,\zeta(2) \, \frac{\left| \epsilon \right|}{\log\left(\left|\epsilon\right|\right)} & \quad \mbox{$d=4$} \\ 
	\frac{\zeta(\frac{d}{2})}{\zeta(\frac{d}{2}-1) + \frac{k_{B}T}{\mu}}\,\left|\epsilon \right| & \quad \mbox{$d>4$} \quad. \\ \end{array} \right.
\end{eqnarray}
The method of steepest descent gives the following expression for the bulk free energy density 
\begin{eqnarray}
\label{bulkdensity<}
\omega_{b}^{>}(T,\mu) = \;\;\;\; \\
-\frac{\mu^2}{2a} - \frac{1}{\beta^2 a}\,\left(\frac{s_{0}^2}{2}-s_{0} \beta \mu \right) - \frac{g_{\frac{d+2}{2}}\left(\exp(s_{0})\right)}{\beta \lambda^d}\, \nonumber \quad,
\end{eqnarray}
where the superscript $>$ denotes the high-temperature case, i.e.,  $\epsilon < 0$.  \\

The bulk correlation length $\xi$ is related to $s_0$ via $\xi\sim |s_0|^{-1/2}$, see Refs.~(\onlinecite{Napiorkowski_11, Napiorkowski_12}). 
From Eq.~(\ref{s0}) we read off the correlation length exponent $\nu(d)$ defined as $\xi \sim |\epsilon |^{-\nu}$ : 
\begin{eqnarray}
\label{nu}
\nu = \left\{ 
\begin{array}{l l}
  {\frac{1}{d-2}} & \quad \mbox{$2<d<4$} \\
	 & \\  
	\frac{1}{2} & \quad \mbox{$d \geq 4$} \\ \end{array} \right. \;,
\end{eqnarray}
where logarithmic corrections to the dominant scaling behavior occur for $d=4$, see Eq.(\ref{s0}).  
The obtained dependence $\nu (d)$, Eq.~(\ref{nu}), is characteristic for the spherical model.\cite{Amit_book,Brankov_00,GB67} Note also that in the case of ideal Bose gas one has $\nu=1/2$.  
The perfect and imperfect Bose gases belong therefore to different bulk universality classes despite the apparent similarities, and the fact that both are exactly solved by the stationary-point approach. 
\subsection{Phase boundary and the low-temperature phase}
At the transition line and in the low-temperature phase (involving the Bose-Einstein condensate), i.e.,  $\epsilon \geq 0$,  the quantity $|s_0|$ vanishes 
in the thermodynamic limit  ($|s_0|\sim V^{-\kappa}, \kappa >0$) irrespective of the choice of $\mu$ and $T$. It is 
now interesting to investigate the character of the singularity treating $V$ as the tuning parameter. 
It turns out that the exponent $\kappa$ relating the vanishing of $s_0$ with the diverging volume depends on $d$ at the phase 
boundary ($\epsilon=0$). The dependence ceases above $d=4$. On the other hand, below the transition temperature we observe no dependence of $\kappa$ on $d$. 
Specifically, for $\epsilon=0$ we find 
\begin{eqnarray}
\label{s01}
|s_{0}(T,\mu_{c})| \approx \left\{ 
\begin{array}{l l}
  {\left[-\frac{\lambda^d}{V}\Gamma(1-\frac{d}{2})^{-1}\right]^{2/d}} & \quad \mbox{$2<d<4$} \\
\left[\frac{\lambda^d}{V}\left(\zeta(\frac{d}{2}-1)+\frac{\lambda^d}{a\beta}\right)^{-1}\right]^{1/2} & \quad \mbox{$d > 4$} \quad. \\ \end{array} \right. 
\end{eqnarray}	
For $d=4$, the logarithmic corrections occur. On the other hand, for $ \epsilon > 0 $ we find
\begin{equation}
|s_{0}(T,\mu)| \approx \frac{\lambda^d}{V}\,\frac{1}{\zeta(d/2)}\,\frac{1}{\epsilon}\;.
\end{equation}
At the phase boundary ($\epsilon =0$) we therefore obtain $\kappa=\frac{2}{d}$ for $2<d<4$, and $\kappa=\frac{1}{2}$ for $d\geq 4$. In the low-temperature 
phase ($\epsilon>0$) we find $\kappa =1$.  \\
In the low-temperature phase the bulk free energy density reads 
\begin{eqnarray}
\label{bulkdensity>}
\omega_{b}^{<}(T,\mu)= -\frac{\mu^2}{2a}  - \frac{1}{\beta \lambda^d}\,\zeta\left(\frac{d+2}{2}\right) \quad.
\end{eqnarray}

We now extract the exponent $\alpha$ characterizing the singular contribution to the bulk free energy 
$\omega_{b}^{sing} \sim |\epsilon|^{2-\alpha}$. From Eqs~(\ref{s0}, \ref{bulkdensity<}) one obtains
\begin{eqnarray}
\label{alfas}
\alpha = \left\{ 
\begin{array}{l l}
  {\frac{d-4}{d-2}} & \quad \mbox{$2<d<4$} \\
	 & \\  
	0 & \quad \mbox{$d \geq 4$} \\ \end{array} \right. \;
\end{eqnarray}
which coincides with the corresponding result for the spherical model.\cite{Amit_book,Brankov_00,GB67} Combined with Eq.~(\ref{nu}) this identifies the bulk universality class of the model in question. 
Thus, although the imperfect Bose gas model is of mean field character, it nevertheless displays nontrivial dependence of the critical indices $\nu$ and $\alpha$ on the system dimensionality. The hyperscaling relation $d \nu = 2-\alpha$ is fulfilled below $d=4$.

\section{Excess surface free energy}
We now focus on a situation, where the system is of finite extent ($D$) in one spatial direction and the $D$-dependence of the free energy gives rise to effective interactions between the boundaries, 
the so-called Casimir forces.\cite{Casimir_48, Chan99, Mostepanienko_97, Krech_94, Kardar_99, Bordag_01, Gambassi_09, Brankov_00, Hucht07, HHG2008,HHGDB2009} These have been a topic of intense studies in very distinct contexts over 
the last decades. The Casimir effect is pronounced in the presence of long-ranged collective excitations, which occur if the system is tuned close to a second-order phase transition, or remains in a phase 
exhibiting generic 
soft modes (for example of Goldstone type). 
There 
are few cases, where interacting systems featuring this effect are susceptible to exact analysis, mainly in two dimensions.\cite{Cardy_86, Maciolek_96, Abraham_06, Jakubczyk_06, Jakubczyk_07, 
Abraham_07, Nowakowski_08, Nowakowski_09, Rudnick_10} Exact results can be also obtained at $d>2$ for the spherical model\cite{Dantchev_96, Chamari_04, Dantchev_04, Dantchev_06} and the related $N\to\infty$ limit of the $O(N)$-symmetric models.\cite{Diehl_12} 

We therefore reconsider the finite, $d$-dimensional IBG and focus on the case $L\gg D$ and also $L$ much larger than any other length present in the system. 
If the thermodynamic state is close to the Bose-Einstein condensation, the bulk correlation length $\xi$ is sizable and by varying $T$ or $\mu$ 
one goes between the situations where $D\ll \xi$ and $D\gg \xi$. For $\epsilon>0$ the system displays 
long-ranged (Goldstone-type) modes and acts coherently also away from the bulk phase boundary. On the other hand, for $\epsilon <0$ there are no long-ranged transverse fluctuations and the bulk correlation length remains finite. The different character of fluctuations at $\epsilon<0$, $\epsilon=0$ and $\epsilon>0$ is reflected in the properties of the Casimir forces acting between the system boundaries separated by $D$. 

The subsequent exact analysis shows how the picture sketched above emerges in the system under study and how the finite-size scaling properties depend on the dimensionality $d$. 

Within the stationary-point approach the excess surface grand canonical free energy per unit area 
\begin{equation}
\label{omega}
 \omega_s(T,D,\mu)=\lim_{L\to\infty}\left[\frac{\Omega(T,L,D,\mu)}{L^{d-1}}-D\,\omega_b(T, \mu)\right]
\end{equation}
is calculated as 
\begin{equation}
\label{omega_2}
 \omega_s(T,\mu,D)=\lim_{L\to\infty}\beta^{-1}D\left[\phi(\bar{s})-\phi_b(s_0)\right]\;,
\end{equation}
where $\phi_b(s_0)$ is evaluated in thermodynamic limit. In the present formulation the analysis amounts to determining $\bar{s}$ from Eq.~(\ref{stat_point_eq}), 
plugging it into 
Eq.~(\ref{omega_2}) and extracting the relevant information in the different scaling regimes. The stationary-point equation can be cast in the form
\begin{eqnarray}
\label{s_equation}
\zeta(d/2)\,\left(-\frac{1}{\beta \mu_{c}} \bar{s}\,+\,\epsilon \right) = g_{d/2}(e^{\bar{s}})\,-\,\zeta(d/2) + \nonumber \\
\frac{2^{3-d/2}}{\pi^{\frac{d-2}{2}}}\,\left(\frac{\lambda}{D}\right)^{d-2}\,
\sum_{n=1}^{\infty} \, \left(\frac{\sigma}{n}\right)^{\frac{d-2}{2}} K_{\frac{d-2}{2}}(n\,\sigma) \, -  \\ 
\frac{{\lambda}^d}{V}\, \sum_{n=0,\pm 1,\pm 2,\dots}^{\infty} \left(1-e^{\frac{\pi \lambda^2 n^2}{D^2}-\bar{s}} \right)^{-1} \nonumber , 
\end{eqnarray}
where $\sigma = 2\pi^{1/2}\frac{D}{\lambda}|\bar{s}|^{1/2}$ and   $K_\rho(z)$ is the modified Bessel function. Eq.~(\ref{s_equation}) holds for $\frac{D}{\lambda}\gg 1$, 
where the discrete sum in $\phi(\bar{s})$, Eq.~(\ref{phi_def}), can be replaced with an integral yielding the function $K_{\frac{d-2}{2}}$. 

We now solve Eq.~(\ref{s_equation}) for $L \rightarrow \infty$ focusing on the limit of small $|\bar{s}|$. This limit corresponds to the system state in the vicinity 
of the bulk phase boundary, or to the bulk low-temperature phase. The following analysis indicates the special role of dimensionalities $d=3$ and $d=4$. 
The special role of $d=3$ is related to the behavior of the sum in Eq.~(\ref{s_equation}) in the limit $\sigma \rightarrow 0$ 
\begin{eqnarray}
\label{sigma1}
\lim_{\sigma \rightarrow 0} \frac{2^{3-d/2}}{\pi^{\frac{d-2}{2}}}\,\left(\frac{\lambda}{D}\right)^{d-2}\,
\sum_{n=1}^{\infty} \, \left(\frac{\sigma}{n}\right)^{\frac{d-2}{2}} K_{\frac{d-2}{2}}(n\,\sigma) = \nonumber \\ 
2 \,\frac{\Gamma(\frac{d-2}{2})}{\pi^{\frac{d-2}{2}}} \, \left(\frac{\lambda}{D}\right)^{d-2} \, \zeta(d-2) \quad.
\end{eqnarray}
The above expression is divergent for $2 < d \leq 3$. On the other hand, it will turn out that for $d>4$ the results become particularly simple. \\ 
To resolve the crossover between $D/\xi\ll 1$ and $D/\xi\gg 1$ for  large  correlation length $\xi\gg \lambda$, we introduce the scaling variable $x$
\begin{eqnarray}
\label{scaling}
x = \left\{
\begin{array}{l l}
  \epsilon \left(\frac{D}{\lambda}\right)^{d-2}  & \quad \mbox{$2<d<4$} \\
                                      &                                     \\
	\epsilon \left(\frac{D}{\lambda}\textsl{}\right)^{2} & \quad \mbox{$d \geq 4$} \quad, \\ 
	\end{array} \right.
\end{eqnarray}
in accord with Eq.~(\ref{nu}). 
We proceed with determining the scaling behavior of the system, in particular the excess surface free energy and the Casimir force.
This has to be done separately for the thermodynamic states corresponding to the bulk low- and high-temperature phases ($\epsilon \geq 0$ and $\epsilon < 0$, respectively).

\subsection{Below bulk condensation temperature, $\epsilon \geq 0$}

For states chosen at or below the bulk condensation temperature, the parameter $\bar{s}$ vanishes for $D/\lambda \to \infty$, alike $s_0$ (compare Sec.~II). At the 
same time the quantity $\sigma = 2\pi^{1/2}\frac{D}{\lambda}|\bar{s}|^{1/2}$ 
may either stay nonzero or vanish, depending on $d$ and $\epsilon$.  
We analyze Eq.~(\ref{s_equation}) via expansion around $\bar{s}=0$. Naturally, the character of the resultant equation is different above and below $d=4$ (see Eq.~(\ref{g_expansion})). It is also clear 
from Eq.~(\ref{sigma1}) that $d=3$ is another borderline dimensionality. We find that above $d=4$ and for any $x>0$ the solution to Eq.~(\ref{s_equation}) behaves asymptotically 
as $V^{-\kappa}$. Therefore the vanishing of $\bar{s}$ is controlled by $L$ rather than $D$. On the other hand, for $x=0$ we find $\bar{s}\sim D^{-(d-2)}$. 
Analysis of the case $d\in (2,4)$ yields a 
solution to Eq.~(\ref{s_equation}) such that the parameter $\sigma$ attains a finite value in the limit $D\gg \lambda$. The particular value of $\sigma$ depends on $x$ and must be determined numerically. 
The relevant equation for $\sigma(x)$ follows from Eq.~(\ref{s_equation}) and reads
\begin{eqnarray}
\label{sigma_eq}
 x \,\zeta\left(\frac{d}{2}\right)\pi^{\frac{d-2}{2}}-\frac{\Gamma(1-\frac{d}{2})}{2^{d-2}}\sigma(x)^{d-2}= \nonumber \\
 2^{3-\frac{d}{2}}\sigma(x)^{\frac{d-2}{2}}\sum_{n=1}^{\infty}n^{-\frac{d-2}{2}}K_{\frac{d-2}{2}}(n\sigma(x))\;.
\end{eqnarray}

The scaling form of the excess surface free energy $\omega_{s}^{<}(T,\mu,D)$ follows from Eq.~(\ref{omega_2}). We obtain 
\begin{eqnarray}
\label{omegas10}
\frac{\omega_{s}^{<}(T,\mu,D)}{k_{B}T} \, = \,-\, \frac{\Delta^{<}(x,d)}{D^{d-1}} \quad.
\end{eqnarray}
For $d\in (2,4)$ the scaling function $\Delta^{<}(x,d)$ takes the form 
\begin{eqnarray}
\label{omegas11}
\Delta^{<}(x,d) = \frac{\zeta(d/2)}{4 \pi} x \sigma(x)^2 + \frac{\Gamma(-d/2)}{2^d \pi^{d/2}}  \sigma(x)^d + \\ 
\frac{2^{2-d/2}}{\pi^{d/2}} \sum_{n=1}^{\infty} \, \left(\frac{\sigma(x)}{n}\right)^{\frac{d}{2}} K_{\frac{d}{2}}(n\,\sigma(x)) \nonumber
\end{eqnarray}
with $\sigma(x)$ obtained as solution of Eq.~(\ref{sigma_eq}). We note that upon shifting the system thermodynamic state away from the critical line  and 
into the low-temperature phase, the scaling function attains the finite value
\begin{equation}
\label{x_limit}
\lim_{x\to\infty} \Delta^{<}(x,d) = 2\,\frac{\Gamma(d/2)}{\pi^{d/2}}\,\zeta(d)\;,
\end{equation}
see Fig. 1.   
On the other hand, for $d \geq 4$ the scaling form of the excess surface free energy $\omega_{s}^{<}(T,\mu,D)$ becomes particularly simple. 
It is given by Eq.~(\ref{omegas10}) with constant scaling function   
\begin{eqnarray}
\label{omegas12}
\Delta^{<}(x,d) = \frac{2\Gamma(\frac{d}{2})}{\pi^{\frac{d}{2}}}\, \zeta(d) \quad.
\end{eqnarray} 
Obviously, Eq.~(\ref{x_limit}) applies also to $d>4$.
We conclude from Eq.~(\ref{omegas12}) that in the low-temperature phase ($\epsilon > 0$) and for $d \geq 4$ the function $\Delta(x,d)$ ceases to 
depend on the scaling variable $x$, see Fig. 1.  
\begin{figure}[h]
\includegraphics[width=8.5cm]{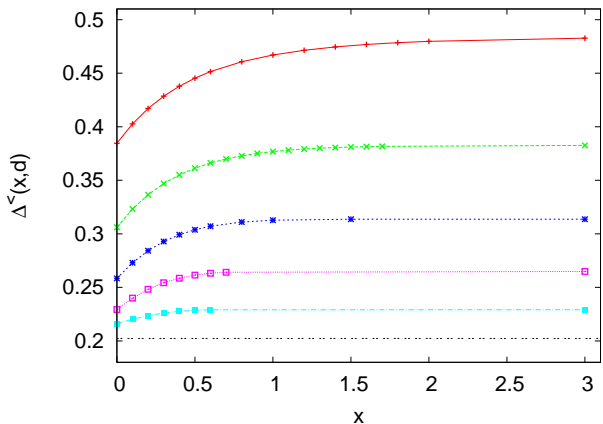}
\caption{(Color online) The scaling function $\Delta^{<}(x,d)$ in the low-temperature phase. The consecutive curves (from top to bottom) correspond to 
$d=2.7$, $d=3.0$, $d=3.3$, $d=3.6$, $d=3.9$ and $d=4.2$. The profiles describe the crossover 
of the excess surface free energy upon moving away from the bulk condensation ($x=0$) into the low-temperature phase ($x\to\infty$),  
where the asymptotic values are given by Eq.~(\ref{x_limit}). For $d>4$ the scaling functions in the low-temperature phase are constant.}
\end{figure}  
We observe that the scaling function is monotonous for $\epsilon >0$. This property is maintained also for $\epsilon < 0$. We also plot the quantities 
$\Delta^{<}(0,d)$ and $\Delta^{<}(x\to\infty,d)$ 
(see Fig.~2).
\begin{figure}[h]
\includegraphics[width=8.5cm]{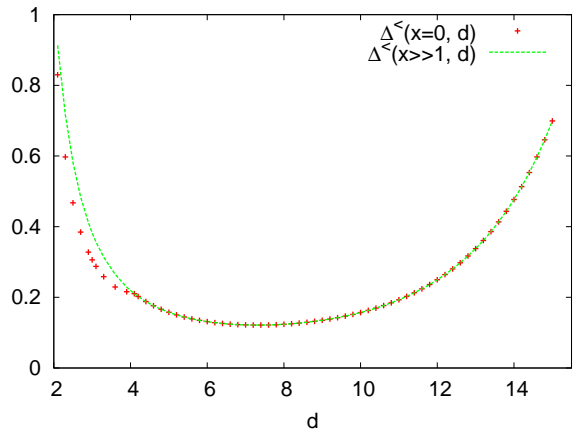}
\caption{(Color online) The amplitudes $\Delta^{<}(0,d)$ and $\Delta^{<}(x\to\infty,d)$ plotted as a function of dimensionality $d$. The two functions merge at $d=4$, 
indicating the evolution of the scaling 
function $\Delta(x,d)$ towards a constant profile as $d\to 4^-$ (compare Fig.~1). The plotted profiles display a minimum at $d\approx 7.4$ and a strong divergence for 
$d\to\infty$.  }   
\end{figure}  
Interestingly, the function $\Delta^<(x,d)$ when projected on any $x>0$, features a minimum at $d\approx 7.4$ and a strong divergence for $d\to\infty$. 

\subsection{Above bulk condensation temperature, $\epsilon < 0$}
For $T>T_c(\mu)$ we look for the solutions of Eq.~(\ref{s_equation}) in the scaling regime corresponding to $x < 0$. In particular, for $D/\lambda \to \infty$ 
we obtain solutions  corresponding to $\bar{s} \to 0$  such that $\sigma$ remains finite. For $2<d<4$ these solutions $\sigma(x)$ fulfill the following equation
\begin{eqnarray}
\label{bars10}
x \zeta(d/2)- \Gamma\left(\frac{2-d}{2}\right) \left(\frac{\sigma}{2 \pi^{1/2}}\right)^{d-2} = \\
+ \, \frac{2^{3-d/2}}{\pi^{d/2-1}} \sum_{n=1}^{\infty} \, \left(\frac{\sigma}{n}\right)^{\frac{d-2}{2}} K_{\frac{d-2}{2}}(n\,\sigma) \quad. \nonumber
\end{eqnarray}
For $d>4$ the  solution of the relevant equation obtained from Eq.~(\ref{s_equation}) has the form
\begin{eqnarray}
\label{bars11}
\sigma =\left(\frac{4 \pi |x|}{\frac{1}{\beta \mu_{c}} + \frac{\zeta(d/2-1)}{\zeta(d/2)}}\right)^{1/2} \quad.
\end{eqnarray}
The analytic expression for the excess surface free energy $\omega_{s}^{>}(T,\mu,D)$ in the scaling regime can be obtained for large negative $x$. It takes the form  
\begin{eqnarray}
\label{omegas20}
\frac{\omega_{s}^{>}(T,\mu,D)}{k_{B}T} \, \approx \,-\, \frac{\Delta^{>}(x,d)}{D^{d}} \quad,
\end{eqnarray}
and 
\begin{eqnarray}
\label{omegas21}
\Delta^{>}(x,d) = \frac{2^{2-d/2}}{\pi^{d/2}} \sum_{n=1}^{\infty} \, \left(\frac{\sigma(x)}{n}\right)^{\frac{d}{2}} K_{\frac{d}{2}}(n\,\sigma(x)) \,,
\end{eqnarray}
where $\sigma(x)$ is obtained as solution of Eq.~(\ref{bars10}) or Eq.~(\ref{bars11}) for $2<d<4$ and $d \geq 4$, respectively. The scaling function decays monotonously 
to zero for increasing $|x|$ which conforms with the 
generically envisaged picture.\cite{Casimir_48, Mostepanienko_97, Krech_94, Kardar_99} It is however worth noting that the monotonous profile of the scaling 
function in the full range of $x$ is specific to the spherical model supplemented with periodic boundary conditions. For example, in the cases of the Ising 
or $XY$ models with periodic boundary conditions, the corresponding scaling functions exhibit a pronounced maximum near $T_c$. So does the 
scaling function obtained recently for the three-dimensional spherical model with free boundary conditions.\cite{Diehl_12}

\section{Casimir forces and critical amplitudes}

The Casimir force $F(T,\mu,D)$ is defined through 
\begin{equation}
F(T,\mu,D) = - \frac{\partial \omega_{s}(T,\mu,D)}{\partial D}\;.
\end{equation}
In the scaling regime it takes the following form 
\begin{eqnarray}
\frac{F(T,\mu,D)}{k_{B}T} = - \frac{\bar{\Delta}(x,d)}{D^d}
\end{eqnarray}
where 
\begin{eqnarray}
\label{scaling4}
\bar{\Delta}(x,d) = \left\{
\begin{array}{l l}
  \left[d-1-(d-2) x \frac{\partial}{\partial x}\right]\Delta(x,d), &\mbox{$2<d<4$} \\
                                      &                                     \\
	\left[(d-1)-2x\frac{\partial}{\partial x}\right]\Delta(x,d),& \mbox{$d \geq 4$} \,. \\ 
	\end{array} \right.
\end{eqnarray}

At the critical line $\mu=\mu_{c}(T)$, i.e., $\epsilon = 0$ the scaling functions $\bar{\Delta}(x,d)$ reduce to the critical amplitudes  
$\bar{\Delta}_{c}(d) = \bar{\Delta}(0,d)=(d-1)\,\Delta(0,d)$
\begin{eqnarray}
\frac{F(T,\mu_{c},D)}{k_{B}T} = - \frac{\bar{\Delta}_{c}(d)}{D^d}
\end{eqnarray}
which can be evaluated on the basis of Eqs(\ref{s_equation},\ref{omegas11},\ref{omegas12}). In the particular, physical case $d=3$ and $x=0$, 
the expressions for the amplitudes can be further simplified. Eq.~(\ref{s_equation}) takes the form
\begin{eqnarray}
\sigma_{3} = -2 \log(1-e^{-\sigma_{3}})
\end{eqnarray}
with the solution $\sigma_{3} =-2 \log(\frac{\sqrt{5}-1}{2})$. After inserting this value to Eqs(\ref{omegas11},\ref{scaling4}) one obtains 
\begin{eqnarray}
\label{omegas31}
\bar{\Delta}_{c}(3)=\frac{2}{\pi}\left[\frac{\sigma_3^3}{6}+\sigma_3 g_{2}(e^{-\sigma_3})+g_{3}(e^{-\sigma_3})\right] \approx  
0.612 . 
\end{eqnarray} 
We note that this value does not coincide with the corresponding result for the noninteracting Bose gas,\cite{Martin_06} contrary to the statement of Ref.~\onlinecite{Napiorkowski_11}, where a relevant 
$D$-dependent term was overlooked in the analyzed equation for $|\bar{s}|$.  
The number obtained in Eq.~(\ref{omegas31}) is exactly twice the value of the critical Casimir amplitude evaluated for the 3-dimensional spherical model.\cite{,Brankov_00,Krech_94} 
The same applies to the asymptotic amplitude for $x\to\infty$, evaluated in Sec.~IV and also to the full profile of the scaling function plotted in Fig.~1. 
The reason of this difference - particularly in view of the results  in Sec. III about the universality class - is not clear to us.  \\

\section{Summary and outlook}

We addressed the issue of the critical bulk properties and Casimir forces of the $d$-dimensional imperfect Bose gas subject to periodic boundary conditions. We started off with performing an exact analysis 
of the critical properties of the system in the thermodynamic limit. The identified universality class is the one of the classical spherical model in $d$ dimensions. We also discussed the onset of 
Bose-Einstein condensation using volume as the tuning parameter at temperature fixed at or below the condensation temperature $T_c(\mu)$. The character of the obtained divergence of the correlation length 
depends on $d$ and whether $T=T_c$, or $T<T_c$. Our results clearly indicate that despite the apparent similarity, the critical properties of the imperfect Bose gas are very distinct to those of the ideal 
Bose gas. In particular, despite being exactly solved by the stationary-point approach, the system exhibits nontrivial dependencies on $d$, as also is the case of the classical spherical model. 

In the subsequent part of the work, we considered the system to be finite in one direction and extracted the excess grand-canonical free energy responsible for the appearance of the Casimir forces. The 
system displays long-ranged Casimir-type interactions for $T\leq T_c$ and the crossover between the $T=T_c$ and $T<T_c$ forms of the excess free energy 
(and consequently also the Casimir force) is governed 
by a universal scaling function, which we compute for varying $d$. The obtained shapes are monotonously increasing for $d\in (2,4)$, while above $d=4$ the scaling function is constant. The particular values 
of the Casimir amplitudes at $T=T_c$ and $T<T_c$ are precisely twice larger than those obtained \cite{Dantchev_96} for the spherical model with periodic boundary conditions. 
The reason of this discrepancy is not clear to us. We also analyzed the Casimir amplitude as a function of $d$. The obtained universal dependence is characteristic to the 
spherical model\cite{Dantchev_96, Dantchev_04} with periodic boundary conditions. It features a minimum around $d\approx 7.4$ and a strong divergence for $d\to\infty$.  


\begin{acknowledgments}
We acknowledge funding by the National Science Centre via 2011/03/B/ST3/02638.
\end{acknowledgments}

\end{document}